\documentclass[aps,prd,twocolumn,floatfix,nofootinbib,preprintnumbers,raggedbottom,10pt]{revtex4-2}
\usepackage[utf8]{inputenc}
\usepackage{caption}
\usepackage{relsize}

\usepackage{amsmath}\usepackage[bottom]{footmisc}
\usepackage{braket}
\usepackage{graphicx}
\usepackage{mwe}
\usepackage{placeins}
\usepackage{framed}
\usepackage{csquotes}
\usepackage{tikz}
\usetikzlibrary{decorations.markings}
\usetikzlibrary{decorations.pathmorphing}
\usepackage{hyperref}
\usepackage{subcaption}
\usepackage{pifont}
\usepackage{setspace}
\usepackage{amsmath, amssymb, amsthm, float, graphicx,amsfonts}
\usepackage{amsthm}
\usepackage{array, boldline, makecell, booktabs}

\theoremstyle{definition}

\hypersetup{hidelinks}
\def\beq{\begin{eqnarray}}\def\eeq{\end{eqnarray}}
\def\be{\begin{equation}}\def\ee{\end{equation}}
\def\bs{\begin{split}}\def\es{\end{split}}

\usepackage{bm}

\usepackage{amstext} %
\usepackage{array}   %
\newcolumntype{C}{>{$}c<{$}}
\newcolumntype{L}{>{$}l<{$}} %

\renewcommand{\thesection}{\Roman{section}}
\allowdisplaybreaks

\begin{document}
\title{Cosmological Magnetic Fluctuations: Positivity Bounds and QCD Contributions}
\author{Ahmadullah Zahed\\
{\it Institute of High Energy Physics, Chinese Academy of Sciences, Beijing 100049, China}\\
{\it China Center of Advanced Science and Technology (CCAST), Beijing 100190, China}\\
\it ICTP, International Centre for Theoretical Physics,\\
\it Strada Costiera 11, 34135, Trieste, Italy\\
zahedahmadullah[AT]gmail.com
}

\begin{abstract}

We study thermal cosmological magnetic fluctuations in two complementary ways. First, we bound the equilibrium magnetic power spectrum in terms of two spectral moments. The proof uses the fluctuation--dissipation relation, spectral positivity, and the convexity and concavity of the thermal kernels. The bounds do not require a model for the spectral shape. They also apply to individual current contributions, using their own moments, and a known frequency threshold strengthens the upper bound. {Second, we calculate a hadronic QCD contribution to cosmological magnetic fluctuations at leading electromagnetic order. We estimate the \(\rho\)-resonance contribution, which dominates the timelike hadronic contribution in our comparison.} Examples from QCD, an electron--positron plasma and a cold collisional plasma test the bounds. The spectra describe fluctuations at the chosen cosmological epoch.

\end{abstract}
\maketitle
\raggedbottom %

\section{Introduction}\label{sec:introduction}
Bootstrap methods use consistency conditions to constrain spectra and interactions, from conformal correlators to scattering amplitudes~\cite{Rattazzi:2008pe,Rodina:2026eji,Chiang:2026nmd,He:2025gws,Bocchia:2026lty}. Positivity bounds similarly constrain effective interactions and response through unitarity and causal dispersion relations~\cite{Adams:2006sv,Lee:2025kgs,Apostolidis:2026mvb,Kehagias:2026opd,Bellazzini:2025bay,Sung:2026nkj,Creminelli:2025rxj}. Here we apply spectral positivity to magnetic fluctuations in an early-universe plasma.

Cosmological magnetic fields affect the microwave background, structure formation and gravitational radiation~\cite{DurrerNeronov2013,Subramanian2016}. Fluctuating electric currents provide a microscopic source of magnetic fluctuations. Response theory and in-in methods relate these fluctuations to the retarded electromagnetic response and current correlations~\cite{Boyanovsky2002formalism,Lemoine1995}. Thermal plasma fluctuations~\cite{Souza2008} and cosmological conductivity~\cite{Baym:1997gq} have been studied in this framework. We ask how much can be inferred from spectral moments when the full frequency dependence is unknown.

This paper has two aims. The first is to bound the equilibrium magnetic {power spectrum} using spectral moments. Starting from the fluctuation--dissipation relation~\cite{Boyanovsky2002formalism}, we apply positivity and the convexity and concavity of the Bose-weighted kernels at fixed spatial momentum. The thermal state need not have Lorentz symmetry~\cite{Creminelli:2025rxj}. {The symmetric upper bound is the Falk--Bruch inequality; see, for example, Ref.~\cite{FalkBruch1969}.} We identify the moments needed for the magnetic bounds and show how to use the same argument for an individual current contribution. A known production threshold strengthens the upper bound.

The second aim is to calculate a hadronic QCD contribution. We use the leading electromagnetic {expression for the magnetic power spectrum from a current} while allowing nonperturbative interactions within the {current--current} correlator. {For this relation between current correlations and the magnetic spectrum, see, for example, Refs.~\cite{Boyanovsky2002formalism,Lemoine1995};} Appendix~\ref{app:reduction} derives its leading term from Weinberg's in-in formula~\cite{Weinberg2005} in our conventions. We take the microscopic current inputs from QCD studies~\cite{RappWambach2000,AyalaThermalRho2012,Atchison:2022yxm,Aarts2015} and estimate the hadronic magnetic contribution with a thermal $\rho$ resonance. We calculate fluctuations at the chosen epoch without assuming that they survive as a relic field.

{We test the bounds on our QCD contributions, an electron--positron plasma and a cold collisional plasma. All three examples satisfy the corresponding bounds. The plasma examples also show that, when the allowed interval is narrow, spectral moments give an accurate estimate of the magnetic fluctuations without knowing the full frequency dependence.}

Section~\ref{sec:analyticity} states the equilibrium magnetic spectrum, and Sec.~\ref{sec:full-bounds} derives the positivity bounds. Section~\ref{sec:qcd-application} introduces the hadronic current contribution and the QCD response models; Sec.~\ref{sec:numerical} gives numerical comparisons. The appendices give the derivations, model inputs and computational details.

\section{Equilibrium magnetic power spectrum}\label{sec:analyticity}
We work at a fixed epoch in a spatially flat Friedmann--Robertson--Walker (FRW) spacetime. The comoving momentum is $k$ and the physical momentum is $k_{\rm phys}=k/a$. We consider the isotropic, parity-even magnetic spectrum. With $B_i=a^{-2}i\epsilon_{i\ell m}k_\ell A_m$, define the magnetic power per transverse polarization by $\langle B_i(\mathbf k)B_j(-\mathbf k)\rangle'=P^T_{ij}P_B(k)$; the prime removes the momentum delta function. Here $P^T_{ij}=\delta_{ij}-\hat k_i\hat k_j$ is the transverse projector.

Over a microscopic correlation time, take $a$ and $T$ as fixed. Let $G^R$ be the transverse retarded photon propagator. If the photon field is in equilibrium, the Kubo--Martin--Schwinger (KMS) relation gives
\begin{equation}
\begin{split}
P_B^{\rm occ,phys}&=\frac{2k_{\rm phys}^2}{\pi}
\int_0^\infty n_B(\omega/T)\operatorname{Im}G^R d\omega,\\
P_B^{\rm sym,phys}&=\frac{k_{\rm phys}^2}{\pi}
\int_0^\infty[1+2n_B(\omega/T)]\operatorname{Im}G^R d\omega.
\end{split}\label{observable:full-powers}
\end{equation}
where $n_B(x)=(e^x-1)^{-1}$ is the Bose factor. The symmetric spectrum ($\mathrm{sym}$) is the half-anticommutator covariance. The occupied spectrum ($\mathrm{occ}$) retains only the Bose occupation weight, with the same finite-temperature response. Equation~(\ref{observable:full-powers}) is exact when $G^R$ is the exact interacting propagator. Boyanovsky et al.~\cite{Boyanovsky2002formalism}, Eqs.~(6.50)--(6.53), derive the equilibrium magnetic spectrum. Appendix~\ref{app:photon-FDT} gives the thermal-trace derivation in our conventions.

The symmetric spectrum includes zero-point weight. We use the occupied prescription, with $2n_B$ at the same finite-temperature response, for the thermal magnetic amplitude. Undamped poles belong to the full photon spectral measure. The free occupied limit is $k_{\rm phys}n_B(k_{\rm phys}/T)$.

Here $\omega=\omega_\eta/a$ is physical frequency. The conversion between comoving and physical spectra is
\begin{equation}
\begin{split}
P_B(k)&=a^{-3}P_B^{\rm phys}(k/a),\\
\Delta_B^2&=\frac{k_{\rm phys}^3P_B^{\rm phys}}{2\pi^2},\qquad
\widehat B=\sqrt{\Delta_B^2}.
\end{split}
\label{DeltaB-def}
\end{equation}
The same conversion holds for a selected current-source contribution. The two transverse polarizations give $d\langle B^2\rangle/d\ln k=2\Delta_B^2$ for the specified power. We evaluate these quantities at a fixed epoch.

\section{Positivity bounds}\label{sec:full-bounds}
For the complete equilibrium magnetic spectrum, define the positive measure and its moments at fixed $k_{\rm phys}$:
\begin{equation}
\begin{split}
d\nu_A^{\rm full}&=\frac2\pi\frac{\operatorname{Im}G^R}{\omega}d\omega,\\
S_A^{\rm full}&=\int d\nu_A^{\rm full},\qquad
C_A^{\rm full}=\int\omega^2d\nu_A^{\rm full}.
\end{split}\label{observable:full-moments}
\end{equation}

Equation~(\ref{observable:full-powers}) becomes
\begin{equation}
\begin{split}
P_B^{\rm occ,phys}&=k_{\rm phys}^2\int h(\omega^2)d\nu_A^{\rm full},\\
P_B^{\rm sym,phys}&=k_{\rm phys}^2\int\phi(\omega^2)d\nu_A^{\rm full},
\end{split}\label{bound-field-FDT}
\end{equation}
where $h(u)=\sqrt u\,n_B(\sqrt u/T)$ and $\phi(u)=\sqrt u[1+2n_B(\sqrt u/T)]/2$. For $T>0$ and $u>0$, we have $h''>0$, $0<h<T$, $\phi''<0$ and $\phi>T$; both $h$ and $\phi$ tend to $T$ at $u=0$. Define
\[
\langle f(u)\rangle=\frac1{S_A^{\rm full}}\int_0^\infty f(\omega^2)d\nu_A^{\rm full},
\qquad \overline u=\frac{C_A^{\rm full}}{S_A^{\rm full}}.
\]
Jensen's inequality~\cite{Jensen1906} and the pointwise comparisons with $T$ give
\[
h(\overline u)\le\langle h\rangle\le T,\qquad
T\le\langle\phi\rangle\le\phi(\overline u).
\]
Multiplying by $k_{\rm phys}^2S_A^{\rm full}$ yields
\begin{widetext}
\begin{equation}
{\begin{aligned}
k_{\rm phys}^2S_A^{\rm full}\Omega_A^{\rm full}n_B(\Omega_A^{\rm full}/T)
&\le P_B^{\rm occ,phys}\le Tk_{\rm phys}^2S_A^{\rm full},\\
Tk_{\rm phys}^2S_A^{\rm full}&\le P_B^{\rm sym,phys}
\le\frac{k_{\rm phys}^2S_A^{\rm full}\Omega_A^{\rm full}}2
[1+2n_B(\Omega_A^{\rm full}/T)]\,,
\end{aligned}}\label{bound-full-field}
\end{equation}
\end{widetext}
where $\Omega_A^{\rm full}=\sqrt{C_A^{\rm full}/S_A^{\rm full}}$.
The occupied lower bound and symmetric upper bound require finite $S_A^{\rm full}>0$ and $C_A^{\rm full}$. The opposite sides require only finite $S_A^{\rm full}$. Appendix~\ref{app:full-bound-proofs} discusses convergence. In particular, polynomial boundedness of $G^R$ alone does not guarantee these moments; any dispersion subtractions and ultraviolet convergence conditions must also be specified. The symmetric upper side is the Falk--Bruch bound~\cite{FalkBruch1969}{; see also Ref.~\cite{Harris1967}.} Appendix~\ref{app:full-bound-proofs} gives the detailed proofs.

The moments must correspond to the spectrum being bounded. For the full field, an unsubtracted dispersion relation gives $S_A^{\rm full}=\operatorname{Re}G^R(0,k_{\rm phys})$ and $C_A^{\rm full}=\lim_{\xi\to\infty}\xi^2G^R(i\xi,k_{\rm phys})$. For a canonically normalized photon field, a regulator that preserves the equal-time commutator gives $C_A^{\rm full}=1$. A Maxwell static response, $G^R(0,k_{\rm phys})=k_{\rm phys}^{-2}$, gives $S_A^{\rm full}=1/k_{\rm phys}^2$. These full-field relations are not imposed on the individual current contributions introduced below.

\subsection{\texorpdfstring{{Contributions from individual currents}}{Contributions from individual currents}}
For a selected coupling-free physical current $j_T$, let $\Pi^R_{j,T}$ be its retarded {current--current correlator} and $\Pi_T^{\rm sym}=\langle\{j_T,j_T\}\rangle/2$ its symmetric correlator. Define $\rho_T=-\operatorname{Im}\Pi^R_{j,T}/\pi\ge0$ for $\omega>0$. The KMS condition gives
\begin{equation}
\Pi^{\rm sym}_T=\pi[1+2n_B(\omega/T)]\rho_T.
\label{main:KMS}
\end{equation}
{At leading electromagnetic order, the stationary magnetic power spectra produced by this current are}
\begin{equation}
\begin{split}
\delta P_B^{\rm occ,phys}&=8\pi\alpha k_{\rm phys}^2
\int_0^\infty n_B(\omega/T)\rho_T|G^R|^2d\omega,\\
\delta P_B^{\rm sym,phys}&=4\pi\alpha k_{\rm phys}^2
\int_0^\infty[1+2n_B(\omega/T)]\rho_T|G^R|^2d\omega.
\end{split}\label{observable:source-powers}
\end{equation}
{The functions in Eq.~(\ref{observable:source-powers}) depend on $(\omega,k_{\rm phys};T)$. We use $\delta P_B$ for the magnetic power spectrum from the chosen current, with spectral density $\rho_T$ and propagation kernel $G^R$; it is not a difference between two plasma models. Appendix~\ref{app:reduction} derives Eq.~(\ref{observable:source-powers}) from the time-domain expression obtained using Weinberg's in-in formula~\cite{Weinberg2005}. A bare Maxwell kernel gives the strict leading electromagnetic result. A dressed kernel resums the specified medium response.}

{These magnetic power spectra have the same thermal kernels as} Eq.~(\ref{observable:full-powers}), with their own positive measure and moments:
\begin{equation}
\begin{split}
d\nu_A&=\frac{8\pi\alpha}{\omega}\rho_T|G^R|^2d\omega,\\
S_A&=\int d\nu_A,\\
C_A&=\int\omega^2d\nu_A,\qquad \Omega_A=\sqrt{C_A/S_A}.
\end{split}\label{bound-field-moments}
\end{equation}
Thus Eq.~(\ref{bound-full-field}) applies with $P_B\to\delta P_B$ and $(S_A^{\rm full},C_A^{\rm full},\Omega_A^{\rm full})\to(S_A,C_A,\Omega_A)$. {The moments belong to this selected contribution; they need not equal the full-field moments.}

{A known frequency threshold gives a stronger upper bound.} For a component supported on $\omega\ge\omega_{\rm th}>0$, let $S_{\rm th},C_{\rm th}$ be its moments and $\Omega_{\rm th}=\sqrt{C_{\rm th}/S_{\rm th}}$. Since $\omega n_B(\omega/T)$ decreases, the threshold sharpens the occupied upper bound:
\begin{equation}
\begin{split}
k_{\rm phys}^2\sqrt{S_{\rm th}C_{\rm th}}n_B(\Omega_{\rm th}/T)
&\le \delta P_{B,\rm th}^{\rm occ,phys}\\
&\le k_{\rm phys}^2S_{\rm th}\omega_{\rm th}n_B(\omega_{\rm th}/T).
\end{split}\label{bound-dressed-threshold}
\end{equation}
The lower endpoint is unchanged.

\section{Magnetic fluctuations from QCD current correlators}\label{sec:qcd-application}\label{sec:hadronic}
We now give a microscopic estimate of the QCD current contribution at leading electromagnetic order, independently of the bounds. We begin with the timelike hadronic spectrum, for which the comparison below supports a dominant $\rho$-resonance contribution, and then include transport in a common response model. Strong interactions remain within the current correlator. Boyanovsky et al.~\cite{Boyanovsky2002formalism}, Sec.~VII, similarly distinguish a general spectral formula from its approximate evaluation. The following subsections specify the hadronic input and photon response.

\subsection{Hadronic current contribution}\label{sec:current-sources}\label{sec:PMF}
{We now apply the current-contribution formula of Sec.~\ref{sec:full-bounds} to the timelike hadronic spectrum.}
For the selected two-pion channel, with pion mass $m_\pi$, the threshold is $\omega_\pi=\sqrt{k_{\rm phys}^2+4m_\pi^2}$. {Its contribution obeys the threshold bound, Eq.~(\ref{bound-dressed-threshold}).} In the leading bare-Maxwell approximation, Eq.~(\ref{observable:source-powers}) reduces to

\begin{equation}
\begin{split}
\delta P_B^{\rm occ,phys}&=8\pi\alpha k_{\rm phys}^2\int_{\omega_\pi}^\infty d\omega\,
\frac{n_B(\omega/T)\rho_T^{\rm had}}{(\omega^2-k_{\rm phys}^2)^2},\\
\rho_T^{\rm had}&=\frac{f_\rho^2m_\rho^3\Gamma_\rho\,\theta(\omega-\omega_\pi)}
{\pi[(\omega^2-k_{\rm phys}^2-m_\rho^2)^2+m_\rho^2\Gamma_\rho^2]}.
\end{split}\label{dPB-spectral}
\end{equation}

This approximation requires a gap from the photon pole and $|\Pi^R_{A,T}|\ll|\omega^2-k_{\rm phys}^2|$ on the selected support. It concerns the positive hadronic contribution; the full response is retained for the transport-inclusive calculation below.

\begin{figure}[t]
\centering
\includegraphics[width=\columnwidth]{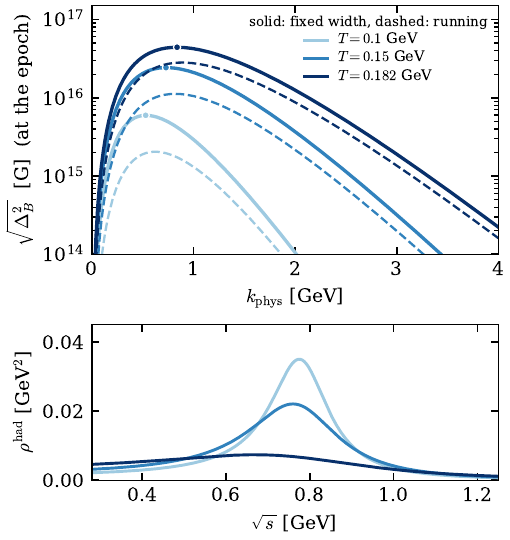}
\caption{Selected hadronic model at $T=0.10,0.15,0.182$~GeV (light to dark): leading occupied magnetic amplitude (top) and transverse $\rho$ spectrum at $k_{\rm phys}=0$ (bottom). Solid/dashed: fixed/running width. This is the fluctuation spectrum at the stated epoch.}
\label{fig:melting}
\end{figure}

We evaluate Eq.~(\ref{dPB-spectral}) with a thermal Breit--Wigner (BW) resonance. Its mass and width use Ref.~\cite{AyalaThermalRho2012}; we retain the vacuum pole coupling as a model choice~\cite{PDG}. This gives a microscopic estimate of the selected hadronic contribution. Here $m_\rho$ and $\Gamma_\rho$ are the resonance mass and width, and $f_\rho$ is its electromagnetic current coupling. The parameter values and running-width test are specified in Appendix~\ref{sm:hadronic-details}. The Maxwell kernel and the Bose factor emphasize the low-mass spectrum. Vacuum $e^+e^-\!\to$ hadrons data show a dominant $\rho$ structure with smaller $\omega$ and $\phi$ resonances~\cite{ParticleDataGroup:2024cfk}, motivating the selected $\rho$ estimate. Figure~\ref{fig:melting} shows its magnetic amplitude and the running-width sensitivity.

At $T=0.15$~GeV and $k_{\rm phys}=0.73$~GeV, keeping the narrow $\omega$ and $\phi$ poles at their measured vacuum values gives occupied powers $1.34\times10^{-7}$ and $5.10\times10^{-8}$~GeV, compared with $1.14\times10^{-5}$~GeV for the thermal $\rho$ model. The $\rho$ term is $98.4\%$ of the sum of these three powers. Separately, reweighting the published vacuum $R$-ratio data in the measured interval $1.5$--$1.99$~GeV gives only $1.50\times10^{-9}$~GeV. Appendix~\ref{sm:hadronic-details} gives the cross-section conversion and inputs without assuming a continuum onset. These checks support a $\rho$-dominated estimate; they do not determine the thermal changes of the other channels.

\subsection{Transport and a common photon response}\label{sec:qcd-resolved}\label{sec:timelike-spacelike}
The current spectrum contains resonance, scattering and transport contributions with different frequency support. Hadronic unitary and scattering cuts occur at finite momentum~\cite{Ghosh:2010hap}; resonance and transport can arise from one hadronic correlator~\cite{Atchison:2022yxm}. Unequal-mass scattering cuts can extend into timelike frequencies, and the zero-momentum conductivity limit is not exclusively spacelike. Quark-phase calculations~\cite{Jackson:2019yao} and transverse lattice information~\cite{Ce:2022fot} supply further QCD input. These studies motivate the spectral and transport inputs used below.

Let $\Pi^R_{A,T}$ be the transverse photon self-energy and $\Gamma_T$ its absorptive part. At frequencies with nonzero absorption,
\begin{equation}
\begin{split}
[G^R]^{-1}&=k_{\rm phys}^2-(\omega+i0)^2+\Pi^R_{A,T},\\
\operatorname{Im}G^R&=|G^R|^2\Gamma_T,\qquad
\Gamma_T=-\operatorname{Im}\Pi^R_{A,T}.
\end{split}\label{main:noise-response}
\end{equation}
For exact $G^R$, $\Pi^R_{A,T}$ is the exact self-energy. In a response model built from the leading current input, the total absorptive density satisfies $\Gamma_T=4\pi^2\alpha\rho_{T,\rm tot}$. If a channel $X$ supplies a positive additive part of this absorption, then
\[
4\pi^2\alpha\rho_{T,X}|G^R|^2
=\frac{\rho_{T,X}}{\rho_{T,\rm tot}}\operatorname{Im}G^R.
\]
This relation selects the contribution of $X$ while keeping the propagation kernel fixed. Correlated sources require the full current covariance, including cross terms (Appendix~\ref{app:qcd-response}).

If all photon modes damp and reach equilibrium through the included currents, summing their stationary source contributions gives the full equilibrium spectrum. An undamped mode retains its initial occupation. Its thermal contribution can instead be obtained by first equilibrating at nonzero width and then taking the width to zero~\cite{Boyanovsky2002formalism,Lemoine1995}. Dividing absorptive densities where $\Gamma_T=0$ does not recover that contribution.

For the model with independent QCD and lepton current fluctuations, both sources use the same dressed photon response. The QCD share has positive measure
\begin{equation}
d\nu_Q=8\pi\alpha\frac{\rho_{T,Q}}{\omega}|G^R|^2d\omega.
\label{QCD-field-measure}
\end{equation}
Using its moments $S_Q,C_Q$ in Eq.~(\ref{bound-full-field}) bounds {the magnetic power spectrum from the QCD current}. Restricting the measure to $\omega\ge\omega_\pi$ gives Eq.~(\ref{bound-dressed-threshold}). We use the following resonance-plus-Drude response, where $Q$ denotes QCD and $\ell$ the lepton current:
\begin{equation}
\begin{split}
[G^R(\omega,k_{\rm phys})]^{-1}={}&k_{\rm phys}^2-(\omega+i0)^2\\
&-\sum_{a=Q,\ell}\frac{i\omega\sigma_a}{1-i\omega\tau_a}+\Pi_\rho^R,\\
\Pi_\rho^R={}&4\pi\alpha\int_{\omega_\pi^2}^\infty du\,
\rho_T^{\rm had}(\sqrt u,k_{\rm phys};T)\\
&\times\left[\frac1{(\omega+i0)^2-u}+\frac1u\right].
\end{split}\label{sm:QCD-response}
\end{equation}
The subtraction sets $G^R(0)=k_{\rm phys}^{-2}$, and the high-frequency limit gives $C_A^{\rm full}=1$. At $T=0.15$~GeV choose $\sigma_Q/T=0.1(4\pi\alpha)(2/3)=0.00611342$, $\sigma_\ell/T=100$, $\tau_Q=T^{-1}$ and $\sigma_\ell/\tau_\ell=4\pi\alpha T^2/9$~\cite{Aarts2015,Baym:1997gq}. Table~\ref{tab:full-checks} compares the full and selected-QCD powers in this model with their respective bounds. In a microscopic calculation, resonance broadening and transport must be obtained consistently from the same current correlator~\cite{Ghosh:2010hap,Atchison:2022yxm}, so that a collision process contributing to both is not counted twice. The additive terms in Eq.~(\ref{sm:QCD-response}) define the phenomenological model tested here.
The resonance density is $\rho_T^{\rm had}$ from Eq.~(\ref{dPB-spectral}). Integration details and source separation are in Appendix~\ref{app:qcd-response}.

\section{Numerical bounds and comparisons}\label{sec:numerical}\label{sec:bound-numbers}
\subsection{\texorpdfstring{{Checks with our QCD contributions}}{Checks with our QCD contributions}}
\subsubsection{Bounds over all frequencies}
At $T=0.15$~GeV, $k_{\rm phys}=0.73$~GeV, Eq.~(\ref{sm:QCD-response}) gives $S_A^{\rm full}=1.8765$~GeV$^{-2}$, $C_A^{\rm full}=1$, as found by spectral integration. Its QCD-sourced part, Eq.~(\ref{QCD-field-measure}), has $S_Q=1.8754$~GeV$^{-2}$ and $C_Q=0.99982$. Table~\ref{tab:full-checks} compares both powers with Eq.~(\ref{bound-full-field}). For the full symmetric field,
\begin{equation}
0.15\le P_B^{\rm sym,phys}=0.370587
\le0.370664\quad\mathrm{GeV}.
\end{equation}
For the full QCD-plus-lepton response, the occupied lower bound lies about $1.5\%$ below the direct integral. The symmetric upper bound lies about $0.021\%$ above the symmetric integral. The occupied upper bound is much looser: it is about 26 times the direct occupied power.
The QCD-source row integrates all frequency support of the resonance-plus-transport model in Eq.~(\ref{sm:QCD-response}). A QCD susceptibility correction can modify its chosen static response~\cite{Bali:2020bcn}.

\begin{table}[t]
\centering\footnotesize\setlength{\tabcolsep}{3pt}

\begin{tabular}{lccc}
\hline\hline
Observable & Lower & Direct & Upper\\\hline
Full QCD+$\ell$ & $0.0056638$ & $0.0057514$ & $0.15$\\
QCD source & $0.0056556$ & $0.0056859$ & $0.14991$\\
\hline\hline
\end{tabular}
\caption{Bounds from Eq.~(\ref{bound-full-field}) and direct occupied powers, all in GeV. QCD rows use Eq.~(\ref{sm:QCD-response}) at $T=0.15$~GeV, $k_{\rm phys}=0.73$~GeV; $\ell$ denotes the charged-lepton model. The QCD-source row uses only $d\nu_Q$ of Eq.~(\ref{QCD-field-measure}) with its own moments.}
\label{tab:full-checks}

\end{table}

\subsubsection{Threshold bounds and the hadronic estimate}
At the same $T,k_{\rm phys}$, the BW integral with a bare kernel gives $\delta P_B^{\rm occ,phys}=1.1388\times10^{-5}$~GeV and $\widehat B=2.43\times10^{16}$~G. Its component moments are $S_{\rm had}=0.0080478$~GeV$^{-2}$ and $C_{\rm had}=0.0065319$. Applying Eq.~(\ref{bound-full-field}) and its threshold refinement, Eq.~(\ref{bound-dressed-threshold}), gives
\[
9.5429\times10^{-6}\le\delta P_B^{\rm occ,phys}
\le1.8401\times10^{-5}\quad\mathrm{GeV}.
\]
The known threshold reduces the bare hadronic upper endpoint by a factor of about 35, from $6.43\times10^{-4}$ to $1.84\times10^{-5}$~GeV. The refined lower and upper endpoints are about $16\%$ below and $62\%$ above the direct estimate, respectively.
Table~\ref{tab:threshold-checks} also tests the dressed selected resonance and the model's complete QCD contribution above threshold. Each row uses the moments of that same positive component; the last row includes its above-threshold transport tail.

\begin{table}[t]
\centering\footnotesize\setlength{\tabcolsep}{3pt}
\begin{tabular}{lccc}
\hline\hline
Selected power & Lower & Direct & Upper\\\hline
Bare BW & $0.95429$ & $1.1388$ & $1.8401$\\
Dressed $\rho$ & $0.92906$ & $1.1110$ & $1.8071$\\
Dressed QCD, $\omega\ge\omega_\pi$ & $0.94728$ & $1.1325$ & $1.8353$\\
\hline\hline
\end{tabular}
\caption{Field-moment bounds with the known two-pion threshold and direct powers, all in $10^{-5}$~GeV. Each interval covers the entire selected contribution.}
\label{tab:threshold-checks}
\end{table}

The result is sensitive to the low-mass spectrum: a density with the alternative numerator in Appendix~\ref{sm:hadronic-details}, normalized to the same pole residue, changes the peak amplitude to about $1.4\times10^{16}$~G at that same epoch. These are local field amplitudes converted using $1~\mathrm G=1.95\times10^{-20}$~GeV$^2$, before any cosmological evolution.

{These comparisons test the bounds on our QCD contributions, with the moments and direct integrals evaluated in the same model. An independent determination of the required moments from QCD is left for future work.}

\subsection{Electron--positron plasma}
We study our bounds on the leading massless $e^+e^-$ response used by Lemoine~\cite{Lemoine1995} and Opher--Opher~\cite{Opher:1996pk}, retaining both the transverse plasmon pole and the spacelike Landau continuum. Direct integration of this spectrum gives $k_{\rm phys}^2S_A^{\rm full}=C_A^{\rm full}=1$ (Appendix~\ref{sm:lemoine}). Equation~(\ref{bound-full-field}) then reads
\[
k_{\rm phys}n_B(k_{\rm phys}/T)\le P_B^{\rm occ,phys}\le T.
\]

We also test the positive current-source contribution itself. In this leading hard-thermal-loop (HTL) approximation, the absorptive density has Landau support $0<\omega<k_{\rm phys}$, while the transverse plasmon is undamped. Selecting that dissipative source in Eq.~(\ref{observable:source-powers}) gives
\[
k_{\rm phys}^2S_{\rm Landau}=0.56106,\qquad
C_{\rm Landau}=0.069182,
\]
and the source interval
\[
0.55117\le\frac{\delta P_{B,\rm Landau}^{\rm occ,phys}}T
=0.55339\le0.56107.
\]
The full occupied power instead has $P_B^{\rm occ,phys}/T=0.96086$; the remaining $0.40747\,T$ comes from the thermally populated plasmon. The source lower endpoint is only $0.40\%$ below its direct value. The full field instead obeys $0.95037\le P_B^{\rm occ,phys}/T\le1$, using its full-field moments. The pole's initial thermal occupation, or equilibration before taking its width to zero, must be specified~\cite{Boyanovsky2002formalism}. The strict dissipative source moments are not the moments of the complete field.

Lemoine's vacuum-subtracted two-polarization spectrum is $\mathcal B_L=2P_B^{\rm sym,phys}-k_{\rm phys}$. Its interval is $2T-k_{\rm phys}\le\mathcal B_L\le2k_{\rm phys}n_B(k_{\rm phys}/T)$. At $k_{\rm phys}=\omega_0=\sqrt{4\pi\alpha}T/3$, direct integration gives
\begin{equation}
\begin{aligned}
1.8990&\le\mathcal B_L/T\le1.9008,\\
\mathcal B_L/T&\simeq1.90076.
\end{aligned}
\label{main:Lemoine-numbers}
\end{equation}
The two temperatures considered in Appendix~\ref{sm:lemoine} give the same dimensionless result. The occupied trace instead has $2P_B^{\rm occ,phys}/T=1.92172$. See that appendix for the derivation and numerical details.

When the spectral moments constrain a sufficiently narrow interval, the bounds determine the equilibrium magnetic fluctuation spectrum accurately without resolving its frequency dependence. Direct plasma calculations above demonstrate such cases.

\subsection{Cold collisional plasma}
Cable--Tajima's cold-plasma response~\cite{CableTajima1992} provides a check with nonzero static spectral weight at $\omega=0$. That weight contributes to $S_A$ and to the power, but not to $C_A$. {Appendix~\ref{sm:cable-tajima} derives the zero-frequency and oscillating weights and gives their numerical comparison with the bounds in Table~\ref{tab:CT-static-weight}.}

The examples test the bounds for resonance, continuum and zero-frequency spectral weight. The QCD powers are estimates from the specified current and response models.

\section*{Acknowledgement}
{I thank Chiara Caprini for inspiration and helpful correspondence.} {I acknowledge} Claude Opus 4.8 (Anthropic) and Codex (Sol 6.1, OpenAI) {for assistance with manuscript polishing and numerical checks.} The author directed this assistance and is responsible for the final content. This work is supported by the National Key R\&D Program of China
(Grant No. 2025YFA1614200) and is supported in part by the National Natural Science Foundation of China under grant No. 12342502. 
I have received support from the European Research Council, grant agreement n. 101039756.

\appendix
\renewcommand{\thesection}{\Alph{section}}
\section{Explicit Derivation of the Magnetic Master Formula}\label{app:reduction}\label{app:BB}
We give an explicit leading-order derivation of the current-source expression in our conventions. With $j_i^c=a^3j_i^{\rm phys}$ and $H_I(\eta)=-e\int d^3x\,A_i j_i^c$, the positive symmetric source term is
\begin{equation}
\begin{split}
\delta P_B^{\rm sym}(k,\eta)={}&\frac{4\pi\alpha k^2}{a^4(\eta)}\int_{\eta_0}^{\eta}d\eta_1d\eta_2\,
G^R_k(\eta,\eta_1)\\
&\times G^R_k(\eta,\eta_2)\Pi_T^{\mathrm{sym}}(\eta_1,\eta_2;k).
\end{split}\label{BB-master}
\end{equation}
Here $\Pi_T^{\rm sym}=\langle\{j_T^c,j_T^c\}\rangle'/2$, with the electromagnetic coupling factored out. Bare kernels $G^R_k=\theta(\eta-\eta')\sin[k(\eta-\eta')]/k$ give the leading $O(\alpha)$ result. Interactions within the source theory need not be perturbative. A dressed kernel resums the specified response, rather than all higher-order terms in the in-in expansion. The derivation below also identifies response acting on the initial field covariance.

\subsection{Quantization of the gauge field}\label{app:quant}
Maxwell theory is conformally invariant in $3+1$ dimensions, so in conformal
time the transverse gauge field obeys the flat-space mode equation
$A_k''+k^2A_k=0$. The Coulomb-gauge mode expansion is
\begin{equation}
\begin{aligned}
A_i(\mathbf x,\eta)=\int\frac{d^3q}{(2\pi)^3}\sum_{\sigma=1,2}
\Bigl[&e_i^\sigma A_q(\eta)\hat a_{\mathbf q\sigma}e^{i\mathbf q\cdot\mathbf x}\\
&+e_i^{\sigma*}A_q^*(\eta)\hat a^\dagger_{\mathbf q\sigma}e^{-i\mathbf q\cdot\mathbf x}\Bigr].
\end{aligned}\label{app-mode}
\end{equation}
The transverse polarizations satisfy $\sum_\sigma e_i^\sigma e_j^{\sigma*}=P^T_{ij}$,
the ladder operators obey $[\hat a_{\mathbf q\sigma},\hat a^\dagger_{\mathbf q'\sigma'}]=(2\pi)^3\delta_{\sigma\sigma'}\delta^{(3)}(\mathbf q-\mathbf q')$, and $A_q(\eta)=e^{-iq\eta}/\sqrt{2q}$.
These are free-field steps. Below $G^R_{A,k}=G^R_{0,k}$; Eq.~(\ref{BB-master}) also allows a specified dressed response. For this derivation define the electromagnetic current $J_i=\sqrt{4\pi\alpha}\,j_i^c$ and its symmetric correlator $\Pi^J_{mp}$; thus $\Pi_T^J=4\pi\alpha\Pi_T^{\rm sym}$ in the main convention. The mean current vanishes in the state considered.

\subsection{The free-field commutator and the Green's function}\label{app:comm}
The free-field commutator is a $c$-number.
\begin{equation}
\begin{aligned}
{}[A_i(\mathbf k,\eta),A_p(\mathbf x_1,\eta_1)]
&=P^T_{ip}e^{-i\mathbf k\cdot\mathbf x_1}\\
&\quad\times\left[-\frac{i}{k}\sin k(\eta-\eta_1)\right],\\
G^R_{A,k}(\eta,\eta_1)&=\theta(\eta-\eta_1)\frac{\sin k(\eta-\eta_1)}{k}.
\end{aligned}\label{app-comm}
\end{equation}
The bracket is purely imaginary; it is the Pauli--Jordan commutator function.
The ordered in-in integral enforces retarded support; $G^R_{A,k}=i\theta[A,A]'$ satisfies $(\partial_\eta^2+k^2)G^R_{A,k}=\delta(\eta-\eta_1)$.

\subsection{Second-order reduction}
Insert the conformal-frame coupling $H_I(\eta_n)=-\int d^3x_n\,A_p(n)J_p(n)$
into Weinberg's second-order formula~\cite{Weinberg2005}, with $Q=A_i(\mathbf k,\eta)A_j(-\mathbf k,\eta)$:
\begin{equation}
\begin{split}
\delta\langle Q\rangle={}&-\int_{\eta_0}^\eta d\eta_2\int_{\eta_0}^{\eta_2}d\eta_1
\int d^3x_1d^3x_2\\
&\times\langle[A_p(1)J_p(1),[A_q(2)J_q(2),Q]]\rangle.
\end{split}\label{app-step1}
\end{equation}
With $C_2=[A_q(2),Q]$, both current channels are
\begin{equation}
[A_pJ_p,[A_qJ_q,Q]]
=\tfrac12[A_p,C_2]\{J_p,J_q\}
+\tfrac12\{A_p,C_2\}[J_p,J_q].
\label{sm:operator-identity}
\end{equation}
The double commutator of the gauge fields is the two-channel $c$-number
\begin{equation}
\begin{aligned}
{}[[A_iA_j,A_q(2)],A_p(1)]={}&[A_i,A_p(1)][A_j,A_q(2)]\\
&+[A_i,A_q(2)][A_j,A_p(1)].
\end{aligned}\label{app-twochannel}
\end{equation}
The two brackets give $i^2=-1$ times two retarded functions, cancelling the $-1$ from the in-in prefactor.
The opposite Fourier phases give $J_p(\mathbf k,\eta_1)J_q(-\mathbf k,\eta_2)$. The two attachments combine the ordered time integrals into an integral over $[\eta_0,\eta]^2$, with
\begin{equation}
\Pi^J_{mp}=\tfrac12\langle\{J_m(\mathbf k,\eta_1),J_p(-\mathbf k,\eta_2)\}\rangle'.
\label{app-PiJ}
\end{equation}
This gives the source term of Eq.~(\ref{BB-master}) with bare kernels. The other channel in Eq.~(\ref{sm:operator-identity}) describes response acting on initial photons. Within the current-source response approximation, the covariance has the form
\begin{equation}
\langle AA\rangle_{\rm sym}=\langle A_{\rm hom}A_{\rm hom}\rangle_{\rm sym}
+G^R\Pi^JG^A.
\label{sm:initial-noise}
\end{equation}
Here $G^A$ is the advanced propagator, and products denote time convolutions and index contractions. The first term propagates the initial photon covariance; the second is generated by the current. Both use the same response. Subtracting free propagation also changes the first term, so that difference need not be positive.

\subsection{Tree-level check}\label{app:tree}
At zeroth order in $H_I$ the gauge correlator is the free one,
$\langle A_iA_j\rangle'_{\rm tree}=P^T_{ij}|A_k|^2=P^T_{ij}/(2k)$. The curl then gives
\begin{equation}
P_B^{\rm tree}(k,\eta)=\frac{k}{2a^4(\eta)}.\label{app-tree}
\end{equation}
This is a check on the mode normalization and on the $B_i=a^{-2}\nabla\times A$ relation.
It is the free vacuum term; an independently populated initial photon state has additional occupation weight.

\subsection{Tensor decomposition and the KMS condition}\label{app:tensorFDT}
At finite temperature the heat-bath four-velocity $u^\mu$ breaks Lorentz invariance, and a spatial rank-two tensor at fixed $\mathbf k$ decomposes into the transverse and longitudinal projectors
\begin{equation}
P^T_{mp}=\delta_{mp}-\hat k_m\hat k_p,\quad P^L_{mp}=\hat k_m\hat k_p,\quad P^T+P^L=\mathbf1.
\label{app-3Dproj}
\end{equation}
Apply this to the spatial retarded correlator and to $\Pi^J_{mp}$. The coefficients depend separately on frequency and spatial momentum; current conservation alone does not imply a single vacuum form factor. In the local physical frame use $\Pi^R_{j,mp}=-i\theta(t)\langle[j_m(t),j_p(0)]\rangle$ and $W^>-W^<=2\pi\rho_{mp}$ for the coupling-free current; $\Pi^J$ below denotes its electromagnetic correlator in this frame. Write $n_B(x)=(e^x-1)^{-1}$. KMS gives $W^<=e^{-\omega/T}W^>$, hence
\begin{equation}
\begin{split}
\Pi_{mp}^{\rm sym}&=\tfrac12(W^>_{mp}+W^<_{mp})
=\pi[1+2n_B(\omega/T)]\rho_{mp},\\
\rho_{T,L}&=-\operatorname{Im}\Pi^R_{j,T,L}/\pi,\qquad
\Pi^J_{mp}=4\pi\alpha\Pi_{mp}^{\rm sym}.
\end{split}\label{app-tensorFDT}
\end{equation}
The retarded-sign convention fixes all factors: $\Pi^R_j(\omega)=\int d\omega'\,\rho(\omega')/(\omega-\omega'+i0)$ up to real subtractions has $\operatorname{Im}\Pi_j^R=-\pi\rho$.

\subsection{Magnetic projection}\label{app:plasma}
The Levi-Civita contraction acts on the projectors as
\begin{equation}
\begin{aligned}
\epsilon_{i\ell m}\epsilon_{jnp}k_\ell k_nP^T_{mp}&=k^2P^T_{ij},\\
\epsilon_{i\ell m}\epsilon_{jnp}k_\ell k_nP^L_{mp}&=0.
\end{aligned}\label{app-curlcontract}
\end{equation}
The curl annihilates the longitudinal channel and converts the transverse projector to $k^2P^T_{ij}$.

The curl therefore selects only the transverse current correlator. The stationary reduction then uses the response identity in Eq.~(\ref{main:noise-response}) and the current KMS relation in Eq.~(\ref{main:KMS}).

\subsection{Equilibrium magnetic spectrum from KMS}\label{app:photon-FDT}
Work at the fixed epoch in physical coordinates, with $\beta=1/T$. For one real transverse polarization write $A_T(t,\mathbf k_{\rm phys})$, and let $\langle O\rangle_\beta=\operatorname{Tr}(e^{-\beta H}O)/\operatorname{Tr}e^{-\beta H}$. Both $H$ and the Heisenberg fields are fully interacting. We suppress the fixed momentum and temperature below; the prime removes the momentum delta function.

Define the two Wightman functions with Fourier convention $e^{i\omega t}$:
\begin{equation}
\begin{split}
W^>_{A,T}(\omega)&=\int_{-\infty}^{\infty}dt\,e^{i\omega t}
\langle A_T(t,\mathbf k_{\rm phys})A_T(0,-\mathbf k_{\rm phys})\rangle'_\beta,\\
W^<_{A,T}(\omega)&=\int_{-\infty}^{\infty}dt\,e^{i\omega t}
\langle A_T(0,-\mathbf k_{\rm phys})A_T(t,\mathbf k_{\rm phys})\rangle'_\beta.
\end{split}\label{app:FDT-Wightman}
\end{equation}
Inserting energy eigenstates in the trace fixes $\omega=E_m-E_n$ through the Fourier integral. Exchanging the states changes the Boltzmann weight from $e^{-\beta E_n}$ to $e^{-\beta E_m}=e^{-\beta\omega}e^{-\beta E_n}$, leaving the matrix-element product unchanged. This gives the KMS relation $W^<_{A,T}=e^{-\beta\omega}W^>_{A,T}$. With our retarded convention,
\begin{equation}
\begin{split}
G^R(t)&=i\theta(t)\langle[A_T(t,\mathbf k_{\rm phys}),
A_T(0,-\mathbf k_{\rm phys})]\rangle'_\beta,\\
2\operatorname{Im}G^R(\omega)&=W^>_{A,T}(\omega)-W^<_{A,T}(\omega).
\end{split}\label{app:FDT-retarded}
\end{equation}
For $\omega>0$, solving these two relations gives
\begin{equation}
\begin{split}
W^>_{A,T}&=2[1+n_B(\omega/T)]\operatorname{Im}G^R,\\
W^<_{A,T}&=2n_B(\omega/T)\operatorname{Im}G^R,\\
\frac{W^>_{A,T}+W^<_{A,T}}2
&=[1+2n_B(\omega/T)]\operatorname{Im}G^R.
\end{split}\label{app:FDT-KMS}
\end{equation}
Thus the symmetric factor follows directly from the thermal trace.

The equal-time covariance is the integral of the symmetric Wightman function over $d\omega/(2\pi)$. The magnetic curls supply $k_{\rm phys}^2$. In the isotropic transverse sector, $\operatorname{Im}G^R$ and $1+2n_B$ are odd in $\omega$, so their product is even. Hence
\begin{equation}
\begin{split}
P_B^{\rm sym,phys}
&=k_{\rm phys}^2\int_{-\infty}^{\infty}\frac{d\omega}{2\pi}
\frac{W^>_{A,T}+W^<_{A,T}}2\\
&=\frac{k_{\rm phys}^2}{\pi}\int_0^\infty d\omega\,
[1+2n_B(\omega/T)]\operatorname{Im}G^R.
\end{split}\label{sm:field-FDT}
\end{equation}
Keeping $2n_B$ at the same finite-temperature $G^R$ gives the occupied prescription in Eq.~(\ref{observable:full-powers}); this is not an interacting zero-temperature subtraction. Boyanovsky et al.~\cite{Boyanovsky2002formalism}, Eqs.~(6.50)--(6.53), give the equilibrium derivation in their conventions, with $S_B=2P_B^{\rm sym,phys}$ here.

\subsection{Conversion to physical variables}\label{app:weyl}
At the fixed epoch $a=a_c$, the physical frequency and momentum obey
\begin{equation}
\omega_\eta=a_c\omega,\qquad\mathbf k=a_c\mathbf k_{\rm phys}.\label{eq:Qrescale}
\end{equation}
The current $J_i=\sqrt{4\pi\alpha}\,a_c^3j_i^{\rm phys}$ gives $\Pi_c^J=a_c^2\Pi_{\rm phys}^J$ after the four-dimensional Fourier Jacobian, while $G_c^R=a_c^{-2}G_{\rm phys}^R$. In the stationary source integral, the factors from $k^2/a_c^4$, $d\omega_\eta$, the two propagators and the current correlator combine as $a_c^{-2}a_c a_c^{-4}a_c^2=a_c^{-3}$. This yields the conversion in Eq.~(\ref{DeltaB-def}) without any assumption about subsequent flux evolution.

Boyanovsky et al.~\cite{Boyanovsky2002formalism}, Eqs.~(6.41)--(6.45), and Calzetta--Kandus~\cite{Calzetta:2001cf}, Eqs.~(25)--(26), give the source--response construction. Boyanovsky's magnetic trace satisfies $S_B=2P_B^{\rm sym,phys}$ in our per-polarization convention. Lemoine~\cite{Lemoine1995}, Eqs.~(13), (16) and (32), and Opher--Opher~\cite{Opher:1996pk}, Eqs.~(5)--(6), give related equilibrium spectral formulations.

\section{Source moments, photon-propagator sum rules and convexity}\label{app:full-bound-proofs}
The positive source measure in Eq.~(\ref{bound-field-moments}) has independent moments $S_A,C_A$. The following dispersion identities apply only to the full photon measure in Eq.~(\ref{observable:full-moments}).

For a passive response with an unsubtracted dispersion relation,
\begin{equation}
G^R(z,k_{\rm phys})=\frac2\pi\int_0^\infty
\frac{\omega\operatorname{Im}G^R(\omega,k_{\rm phys})}{\omega^2-z^2}d\omega,
\qquad\operatorname{Im}z>0.
\end{equation}
At $z=0$ this gives $S_A^{\rm full}=\operatorname{Re}G^R(0,k_{\rm phys})$. At $z=i\xi$, monotone convergence gives

\begin{equation}
C_A^{\rm full}=\frac2\pi\int_0^\infty\omega\operatorname{Im}G^R d\omega
=\lim_{\xi\to\infty}\xi^2G^R(i\xi,k_{\rm phys}).
\end{equation}

For a canonically normalized photon coordinate, a regulator preserving the commutator gives $C_A^{\rm full}=\langle[A,[H,A]]\rangle=1$. A truncated response or a different field normalization need not obey this relation. At fixed $k_{\rm phys}>0$, an Ohmic infrared response has $\operatorname{Im}G^R\propto\omega$, so both moments converge there. Polynomial boundedness by itself guarantees neither moment. If the spectral tail is $\operatorname{Im}G^R\sim\omega^{-p}$, its integrals require $p>0$ for $S_A^{\rm full}$ and $p>2$ for $C_A^{\rm full}$. Dispersion subtraction terms must be specified separately when an unsubtracted representation is unavailable. If only the static moment exists, the occupied upper and symmetric lower sides remain valid. The other two bounds also need finite $C_A$. A positive part of the spectrum with finite moments can still give a lower bound on the full occupied power even if the full second moment diverges. Subtracting a zero-temperature spectrum can produce a signed measure, to which this positivity argument does not directly apply.

For example, an inclusive QCD current continuum with $\rho_T\propto\omega^2$ and a bare Maxwell kernel $|G^R|^2\propto\omega^{-4}$ gives $C_A\propto\int^\infty d\omega/\omega$. The finite moments of the selected BW model must therefore not be assigned to the entire inclusive hadronic spectrum. The occupied power remains ultraviolet convergent because of the Bose factor, and its static-moment upper bound remains available when $S_A$ is finite.

\subsection{Inequalities from convexity and concavity of $h$ and $\phi$}
\begin{figure}[t]
\centering
\includegraphics[width=0.49\columnwidth]{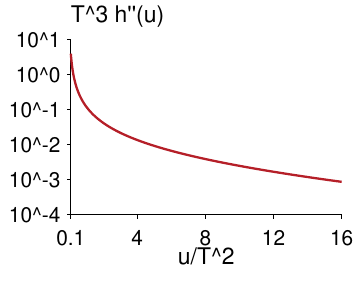}
\caption{The dimensionless curvature $T^3h''(u)$ as a function of $u/T^2$, on a logarithmic vertical scale. The Taylor-series argument establishes its positivity for all $u>0$.}
\label{fig:kernel-curvature}
\end{figure}

Starting directly from $h(u)=\sqrt u/(e^{\sqrt u/T}-1)$ and writing $x=\sqrt u/T$, differentiation gives
\begin{equation}
h''(u)=\frac{e^{2x}(x^2-x-1)+e^x(x^2+x+2)-1}
{4T^3x^3(e^x-1)^3}.
\label{sm:occupied-convexity}
\end{equation}
The denominator is positive for $u,T>0$. The Taylor expansion of its numerator is
\[
\frac{x^3}{2}+\frac{3x^4}{4}+\frac{5x^5}{8}+\frac{131x^6}{360}+\cdots.
\]
More generally, the coefficient of $x^n/n!$ in this numerator is $2^{n-2}(n-4)(n+1)+n^2+2$ for $n\ge3$. It equals $3$ at $n=3$ and is positive for every $n\ge4$; the coefficients through $x^2$ vanish. The exponential series converges for all real $x$, so this proves $h''(u)>0$ for every $u>0$. Also $0<h(u)<T$ follows from $e^x-1>x$. Figure~\ref{fig:kernel-curvature} illustrates the curvature.

The partial-fraction expansion of the symmetric kernel is
\begin{equation}
\begin{split}
\phi(u)&=T+2T\sum_{n=1}^{\infty}\frac{u}{u+(2\pi nT)^2},\\
\phi''(u)&=-4T\sum_{n=1}^{\infty}\frac{(2\pi nT)^2}{[u+(2\pi nT)^2]^3}<0.
\end{split}\label{sm:symmetric-concavity}
\end{equation}
The first line also gives $\phi(u)>T$. Both kernels extend continuously to $h(0)=\phi(0)=T$.

\begin{figure}[t]
\centering
\includegraphics[width=\columnwidth]{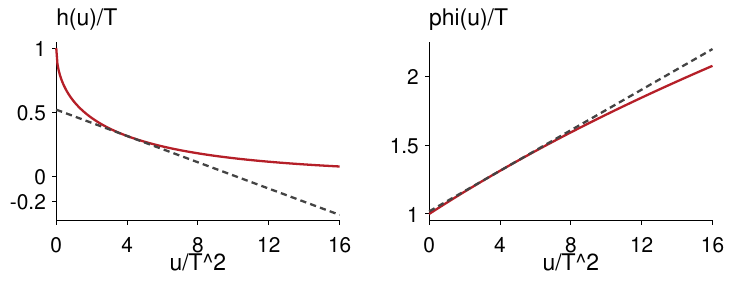}
\caption{Supporting tangents at $\bar u=4T^2$. The solid curves show $h(u)/T$ (left) and $\phi(u)/T$ (right) as functions of $u/T^2$; dashed lines are their tangents. Convexity puts $h$ above its tangent, while concavity puts $\phi$ below it.}
\label{fig:kernel-tangents}
\end{figure}

For either positive measure used in the paper, write its moments as $S_A,C_A$ and set $\bar u=C_A/S_A$. Convexity and concavity give the supporting-tangent inequalities
\begin{equation}
\begin{split}
h(u)&\ge h(\bar u)+h'(\bar u)(u-\bar u),\\
\phi(u)&\le\phi(\bar u)+\phi'(\bar u)(u-\bar u).
\end{split}\label{app:tangent-bounds}
\end{equation}
See figure \ref{fig:kernel-tangents} for illustration. Integrating with the probability measure $d\nu_A/S_A$ on $0\le\omega<\infty$, the linear terms vanish because
\[
\frac1{S_A}\int(\omega^2-\bar u)d\nu_A
=\frac{C_A}{S_A}-\bar u=0.
\]
Thus $\langle h\rangle\ge h(\bar u)$ and $\langle\phi\rangle\le\phi(\bar u)$. Combining these with $h\le T\le\phi$ and multiplying by $k_{\rm phys}^2S_A$ proves Eq.~(\ref{bound-full-field}) and its selected-source version. The same argument applies to a positive restriction of the spectrum, with the moments of that restriction.

For the full measure of one real photon coordinate the Duhamel variables are $b=TS_A^{\rm full}$ and $c=C_A^{\rm full}/T$. The symmetric upper side is the Falk--Bruch estimate $\langle A^2\rangle\le\sqrt{bc}[1+2n_B(\sqrt{c/b})]/2$~\cite{FalkBruch1969}. Related thermal bounds are due to Harris~\cite{Harris1967}; applications include quantum-spin ordering~\cite{DysonLiebSimon1978} and quantum spin glasses~\cite{Leschke:2021pre}.

\subsection{Related model independent bootstrap and positivity methods}\label{app:related-methods}
S-matrix and conformal bootstrap methods constrain amplitudes and correlators through consistency conditions~\cite{Rattazzi:2008pe,Bocchia:2026lty,Chowdhury:2026bfz,Wan:2026zss,Chiang:2026ubv,Rodina:2026eji,Backus:2026qdp,Guo:2026xyl,Ferretti:2026iij,Kangshabanik:2026onk,Ghosh:2026dse,Shao:2026akl,deRham:2026lvc,Das:2026ckf,Xu:2026kix,Khansari:2026drs,Wan:2026pjq,EliasMiro:2026utl,Ghosh:2026xnp,Ghosh:2026gku,Chiang:2026nmd,Gumus:2026mhb,Saha:2026ftv,Gumus:2025hwq,Huang:2025icl,Laliberte:2025xvk,Benedetti:2025nzp,Antunes:2025vvl,deRham:2025vaq,Bhat:2025zex,He:2025gws,Berman:2024eid,Guerrieri:2024jkn,Banerjee:2026zhx}. Related positivity studies constrain interactions and causal response, and examine cosmological spectral representations~\cite{Adams:2006sv,Juliano:2026usx,Dong:2026rbc,Cheung:2026wrz,Sung:2026nkj,Apostolidis:2026mvb,Bu:2026krx,Kremminger:2026edr,Kehagias:2026opd,Desai:2026lyt,Peng:2026ztp,Chang:2026ztn,Fernandez:2026fby,Choudhury:2026sah,Deb:2026enc,Lee:2025kgs,Bellazzini:2025bay,deRham:2025htd,Dong:2025dpy,Liu:2025meq,Bellazzini:2025shd,Peng:2025klv,Serra:2024tmz,Chakraborty:2024ciu,Aoude:2024xpx,McPeak:2023wmq,Bal:2026kzn}.

In QCD, spectral moments enter finite-temperature vector-current sum rules~\cite{Gubler:2017qbs}. H\"older inequalities~\cite{Holder1889} have been combined with QCD sum rules to bound the hadronic muon anomaly~\cite{Li:2024frm}; a complementary bootstrap calculation uses unitarity, crossing and finite-energy sum rules~\cite{Zahed2024gmu}. Their kernels and subtraction conventions must be matched before using such moment input in the magnetic bounds.

\section{Hadronic spectral input and high-mass check}\label{sm:hadronic-details}
Equation~(\ref{dPB-spectral}) uses $m_\pi=0.13957$, $m_{\rho,0}=0.775$, $\Gamma_{\rho,0}=0.149$ and $f_\rho=0.156$ in GeV~\cite{PDG}. The fits of Ref.~\cite{AyalaThermalRho2012} are
\[
\begin{aligned}
m_\rho(T)&=m_{\rho,0}[1-(T/0.222~\mathrm{GeV})^{10}],\\
\Gamma_\rho(T)&=\frac{\Gamma_{\rho,0}}{1-(T/0.197~\mathrm{GeV})^3}.
\end{aligned}
\]
We hold the pole coupling $f_\rho$ at its vacuum value. At the main benchmark, $T=0.15$~GeV and $k_{\rm phys}=0.73$~GeV, these inputs give $m_\rho=0.759630$ and $\Gamma_\rho=0.266759$~GeV. The moments follow by inserting the bare response into Eq.~(\ref{bound-field-moments}); their numerical values and the powers are reported in the main text. The envelope $\rho_T^{\rm had}\le f_\rho^2m_\rho/(\pi\Gamma_\rho)$ bounds the occupied power above $12$~GeV by $2.84\times10^{-43}$~GeV at this benchmark, a negligible tail.

For comparison, we also use the numerator $f_\rho^2(\omega^2-k_{\rm phys}^2)m_\rho\Gamma_\rho/\pi$, with the same pole residue. For the running-width curves in Fig.~\ref{fig:melting}, set $M^2=\omega^2-k_{\rm phys}^2$ within the model and use
\[
\Gamma(M)=\Gamma_\rho\frac{m_\rho}{M}
\left(\frac{M^2-4m_\pi^2}{m_\rho^2-4m_\pi^2}\right)^{3/2}\theta(M-2m_\pi),
\]
replacing $m_\rho\Gamma_\rho$ by $M\Gamma(M)$~\cite{Gounaris:1968mw}. This is a sensitivity test of the low-mass input.

\subsection{Vacuum cross-section and resonance comparison}
For vacuum input, $R(M^2)=\sigma(e^+e^-\!\to\mathrm{hadrons})/[4\pi\alpha^2/(3M^2)]$ gives $\rho_T^{\rm vac}=M^2R(M^2)/(12\pi^2)$, where $M^2=\omega^2-k_{\rm phys}^2>0$. With $E_M=\sqrt{k_{\rm phys}^2+M^2}$, Eq.~(\ref{dPB-spectral}) becomes
\begin{equation}
\delta P_{B,R}^{\rm occ,phys}
=\frac{2\alpha k_{\rm phys}^2}{3\pi}
\int_{M_1}^{M_2}\frac{R(M^2)n_B(E_M/T)}{M E_M}\,dM.
\label{app:R-data-power}
\end{equation}
We linearly interpolate the 874 distinct central values in the PDG EXCLSUM 19 tabulation on $0.30$--$1.99$~GeV, averaging its two repeated-energy pairs~\cite{ParticleDataGroup:2024cfk}.\footnote{See the \href{https://pdg.lbl.gov/2025/hadronic-xsections/hadron.html}{PDG hadronic cross-section collection}.} At the main benchmark the covered-range power is $2.23\times10^{-6}$~GeV. The higher-mass portion quoted in Sec.~\ref{sec:hadronic} uses these measured values; it does not assume a perturbative continuum threshold.

For a narrow vector state, $\langle0|j_\mu|V\rangle=f_VM_V\epsilon_\mu$ and $f_V^2=3M_V\Gamma_{Vee}/(4\pi\alpha^2)$. Its density $f_V^2M_V^2\delta(\omega^2-k_{\rm phys}^2-M_V^2)$ gives
\begin{equation}
\begin{split}
\delta P_{B,V}^{\rm occ,phys}
&=\frac{4\pi\alpha k_{\rm phys}^2f_V^2}{M_V^2E_V}n_B(E_V/T),\\
E_V^2&=k_{\rm phys}^2+M_V^2.
\end{split}
\label{app:narrow-vector-power}
\end{equation}
The 2025 PDG inputs are $M_\omega=0.78266$~GeV, $\Gamma_{\omega ee}=0.643188$~keV (from $8.68$~MeV times $0.741\times10^{-4}$), $M_\phi=1.019460$~GeV and $\Gamma_{\phi ee}=1.27$~keV~\cite{ParticleDataGroup:2024cfk}. They give the narrow-pole powers used with the thermal $\rho$ model in the main text. The inclusive $R$ spectrum is a separate comparison and already contains these resonances.

\section{Electron--positron response and Lemoine comparison}\label{sm:lemoine}
For the leading massless pair response~\cite{Lemoine1995}, let $v=\omega/k_{\rm phys}$ and $m_D^2=4\pi\alpha T^2/3$. Related cosmological thermal-fluctuation applications appear in Ref.~\cite{Souza2008}. The transverse propagator is
\begin{equation}
\begin{split}
G^R&=[k_{\rm phys}^2-(\omega+i0)^2+\Pi_T^R]^{-1},\\
\Pi_T^R&=\frac{m_D^2}{2}\left[v^2+\frac{v(1-v^2)}2
\log\frac{v+1+i0}{v-1+i0}\right].
\end{split}
\end{equation}
It has Landau weight on $0<\omega<k_{\rm phys}$ and a transverse plasmon pole at $\omega_T>k_{\rm phys}$, with $Z_T=[1-\partial\Pi_T^R/\partial\omega^2]^{-1}_{\omega_T}$. Equation~(\ref{sm:field-FDT}) gives
\begin{equation}
\begin{split}
P_B^{\rm occ,phys}=k_{\rm phys}^2\biggl[&\frac2\pi\int_0^{k_{\rm phys}}
n_B(\omega/T)\operatorname{Im}G^R d\omega\\
&+\frac{Z_T}{\omega_T}n_B(\omega_T/T)\biggr].
\end{split}\label{sm:pair-integral}
\end{equation}
For the symmetric power replace $2n_B(\omega/T)$ by $1+2n_B(\omega/T)$ in the continuum and $n_B(\omega_T/T)$ by $\tfrac12+n_B(\omega_T/T)$ at the pole. At $k_{\rm phys}=\omega_0=\sqrt{4\pi\alpha}T/3$, $\omega_T/k_{\rm phys}=1.456235$ and $Z_T=0.930818$. The pole and continuum together obey the complete-field moment relations. At this fixed $k_{\rm phys}/T$, powers scale linearly with $T$; the $0.20$-GeV result is twenty times the $0.01$-GeV result.

Lemoine's two-polarization vacuum subtraction is $\mathcal B_L=2P_B^{\rm sym,phys}-k_{\rm phys}$~\cite{Lemoine1995}. The symmetric bound therefore gives
\begin{equation}
2T-k_{\rm phys}\le\mathcal B_L\le2k_{\rm phys}n_B(k_{\rm phys}/T).
\label{bound-Lemoine-subtracted}
\end{equation}
Substitution at $k_{\rm phys}=\omega_0$ gives explicitly
\begin{equation}
1.899059293\le\mathcal B_L/T=1.900756888
\le1.900757176.
\label{sm:Lemoine-numbers}
\end{equation}
This is the same quantity and normalization as Eq.~(\ref{main:Lemoine-numbers}), not the occupied one-polarization power.

For the strict HTL dissipative source, restrict Eq.~(\ref{bound-field-moments}) to $0<\omega<k_{\rm phys}$ using $\rho_T=(-\operatorname{Im}\Pi_T^R)/(4\pi^2\alpha)$. Its moments are
\[
\begin{aligned}
S_{\rm Landau}&=\frac2\pi\int_0^{k_{\rm phys}}\frac{\operatorname{Im}G^R}{\omega}d\omega,\\
C_{\rm Landau}&=\frac2\pi\int_0^{k_{\rm phys}}\omega\operatorname{Im}G^R d\omega.
\end{aligned}
\]
The omitted positive plasmon measure has $S_{\rm pole}=Z_T/\omega_T^2$ and $C_{\rm pole}=Z_T$. Adding them gives $S_A^{\rm full}$ and $C_A^{\rm full}$, while adding its occupied power gives Eq.~(\ref{sm:pair-integral}). The continuum integral reproduces the selected Landau-source result in the main text. An exactly undamped pole cannot be inferred from the regular absorptive density alone. Giving it a nonzero width, equilibrating, and then taking the width to zero instead includes its thermal atom in the stationary total source~\cite{Boyanovsky2002formalism}, Eqs.~(6.58)--(6.63). This differs from selecting only the strict dissipative Landau component.

\section{Selected QCD current contributions}\label{app:observable-audit}\label{app:qcd-response}\label{app:survival}
For the response in Eq.~(\ref{sm:QCD-response}), the coupling-free Drude densities are
\[
\rho_{T,a}^{\rm Drude}(\omega)=
\frac{\omega\sigma_a}{4\pi^2\alpha(1+\omega^2\tau_a^2)},\qquad a=Q,\ell.
\]
Use $\rho_{T,Q}=\rho_T^{\rm had}+\rho_{T,Q}^{\rm Drude}$ and $\rho_{T,\ell}=\rho_{T,\ell}^{\rm Drude}$ in Eq.~(\ref{observable:source-powers}), with the same $G^R$ in both shares. The additive model assumes independent currents, $\Pi_{Q\ell}^{\rm sym}=\Pi_{\ell Q}^{\rm sym}=0$; correlated sources require these cross terms.

Removing QCD instead changes $G^R$. At the main benchmark, recomputing the lepton-only response gives an occupied-power difference $3.02719\times10^{-5}$~GeV. This differs from the positive {magnetic power spectrum from the QCD current}, $0.0056859$~GeV in Table~\ref{tab:full-checks}, which keeps the common response fixed.

\section{Cold-plasma zero-frequency weight}\label{sm:cable-tajima}
The isotropic collisional response of Cable--Tajima~\cite{CableTajima1992} is
\begin{equation}
G^R(z,k_{\rm phys})=
\left[k_{\rm phys}^2-z^2+\frac{\omega_p^2z}{z+i\eta}\right]^{-1},
\qquad\eta>0.
\label{sm:CT-response}
\end{equation}
Its static and high-frequency limits give $S_A^{\rm full}=k_{\rm phys}^{-2}$ and $C_A^{\rm full}=1$. Equilibrate at finite $\eta$ before taking $\eta\to0^+$; with $E^2=k_{\rm phys}^2+\omega_p^2$, the positive moment measure becomes
\begin{equation}
d\nu_A^{\rm full}\to
\frac{\omega_p^2}{k_{\rm phys}^2E^2}\delta_0(d\omega)
+\frac1{E^2}\delta_E(d\omega).
\label{sm:CT-atoms}
\end{equation}
Here $\delta_\omega$ is a unit point measure. The zero-frequency part contributes $\omega_p^2/(k_{\rm phys}^2E^2)$ to $S_A$ and zero to $C_A$. Since $h(0)=\phi(0)=T$, it contributes $T\omega_p^2/E^2$ to either power. The occupied result is therefore $P_{B,\rm CT}^{\rm occ,phys}=T\omega_p^2/E^2+(k_{\rm phys}^2/E)n_B(E/T)$, their Eq.~(58) in our per-polarization convention. Table~\ref{tab:CT-static-weight} gives the two pieces at $k_{\rm phys}=\omega_p=T$. Discarding the zero-frequency part would remove half the static moment and $0.5T$ of the power while leaving $C_A$ unchanged.

\begin{table}[htbp]
\centering\footnotesize\setlength{\tabcolsep}{4pt}
\begingroup
\begin{tabular}{lrrr}
\hline\hline
Part & $k_{\rm phys}^2S_A$ & $C_A$ & Occupied power/$T$\\\hline
Zero frequency & $0.5$ & $0$ & $0.50000$\\
Frequency $E$ & $0.5$ & $1$ & $0.22713$\\
Complete field & $1$ & $1$ & $0.72713$\\
\multicolumn{3}{l}{Complete-field bounds} & $[0.58197,1]$\\
\hline\hline
\end{tabular}
\caption{{Cold plasma at $k_{\rm phys}=\omega_p=T$, where $\omega_p$ is the plasma frequency and $E=\sqrt{k_{\rm phys}^2+\omega_p^2}$. The two components sum to the complete field. The final row gives its occupied-power interval from Eq.~(\ref{bound-full-field}).}}
\label{tab:CT-static-weight}
\endgroup
\end{table}

\section{\texorpdfstring{Smoothed spectra and hadronic limits}{Smoothed spectra and hadronic limits}}\label{app:smoothed-limits}
\subsection{Gaussian-smoothed pair-plasma comparison}
Here $l$ is a physical smoothing length. Filtering each magnetic-field mode by $e^{-k_{\rm phys}^2l^2/2}$, define
\begin{equation}
\begin{split}
B_l^2&\equiv\int_0^\infty\frac{dk_{\rm phys}\,k_{\rm phys}^2}{2\pi^2}
 e^{-k_{\rm phys}^2l^2}\mathcal B_L(k_{\rm phys}),\\
\mathcal B_L(k_{\rm phys})&=2P_B^{\rm sym,phys}(k_{\rm phys};T)-k_{\rm phys}.
\end{split}\label{app:Bl-definition}
\end{equation}
The factor 2 traces over both transverse polarizations. The subtraction removes the free zero-temperature Maxwell variance, whose per-polarization power is $k_{\rm phys}/2$, and retains the medium-dependent change in zero-point weight. Thus $B_l$ is the square root of this subtracted, smoothed variance, in Lemoine's convention~\cite{Lemoine1995}; it is not the occupied variance obtained by replacing the symmetric weight by $2n_B$ at the same finite-temperature response.
Use $\omega_0=\sqrt{4\pi\alpha}T/3$ and $l=2\pi/\omega_0$.
Assume $k_{\rm phys}^2S_A^{\rm full}=C_A^{\rm full}=1$ over the momenta contributing to the integral. Integrating Eq.~(\ref{bound-Lemoine-subtracted}) with the weight in Eq.~(\ref{app:Bl-definition}) gives
\begin{equation}
\begin{split}
\frac{T}{4\pi^{3/2}l^3}-\frac1{4\pi^2l^4}
&\le B_l^2,\\
B_l^2&\le\frac1{\pi^2}\int_0^\infty dk_{\rm phys}\,k_{\rm phys}^3\\
&\qquad\times e^{-k_{\rm phys}^2l^2}n_B(k_{\rm phys}/T).
\end{split}
\end{equation}
At $l=2\pi/\omega_0$, $B_l$ lies in $[2.2025488,2.2025847]\times10^{12}$~G at 10~MeV and $[8.8101952,8.8103386]\times10^{14}$~G at 200~MeV. These agree with Lemoine's rounded Table~I estimates~\cite{Lemoine1995}. The comparison uses the specified soft pair response and Lemoine's leading classical estimate.

\subsection{Low-momentum spectrum}\label{app:dPB-analytic}
At small physical momentum, the leading term factorizes into $k_{\rm phys}^2$ and an integral independent of momentum.
With the spectral convention above, the physical-frame formula is
\begin{equation}
\begin{split}
\delta P_B^{\rm occ,phys}(k_{\rm phys})&=4\pi\alpha k_{\rm phys}^2\widetilde K(T)
+O(k_{\rm phys}^4),\\
\widetilde K(T)&=2\int_{2m_\pi}^\infty d\omega\,
\frac{n_B(\omega/T)\rho_T^{\rm had}(\omega,0;T)}{\omega^4}.
\end{split}\label{app-dPB-lowk}
\end{equation}
It reproduces the scaling $\delta P_B\propto k_{\rm phys}^2$, hence $\widehat B\propto k_{\rm phys}^{5/2}$.
The physical spectrum has no scale factor; its comoving power is $a_c^{-3}$ times this value. This conversion does not assume that the pole dominates the low-mass spectral tail. The expansion requires small momentum relative to the scales controlling the integral, including the two-pion threshold. At $k_{\rm phys}=0.05$~GeV, its excess over the full BW integral is $5.23\%$, $3.93\%$ and $3.64\%$ at $T=0.10,0.15,0.182$~GeV.

\subsection{Location of the spectral peak}\label{app:peak}
Evaluating the thermal factor at the resonance, $\omega_*\equiv\sqrt{k_{\rm phys}^2+m_\rho^2(T)}$, gives
\begin{equation}
\widehat B^2(k_{\rm phys})\propto\frac{k_{\rm phys}^5n_B(\omega_*/T)}{\omega_*}.\label{app-Bpeak-model}
\end{equation}
Differentiating this assumed shape gives $5\omega_*^2=k_{\rm phys}^2[1+(\omega_*/T)(1+n_B(\omega_*/T))]$. In its Wien limit, the stationarity condition and cubic are
\begin{equation}
\begin{aligned}
5(k_{\rm phys}^2+m_\rho^2)&=k_{\rm phys}^2(1+\omega_*/T),\\
\omega_*^3-4T\omega_*^2-m_\rho^2\omega_*-Tm_\rho^2&=0.
\end{aligned}\label{app-peak-cubic}
\end{equation}
Dropping the $Tm_\rho^2$ term gives
\begin{equation}
k_{\rm peak}\simeq2\sqrt{T\left(2T+\sqrt{4T^2+m_\rho^2(T)}\right)}.\label{app-kpeak}
\end{equation}
At $T=0.10,0.15,0.182$~GeV, the full BW peaks are $0.53164,0.73098,0.83716$~GeV; the simplified estimate gives $0.63250,0.81856,0.90513$~GeV. Its relative errors are about $19\%$, $12\%$ and $8\%$, respectively. The peak depends on $T$, $m_\rho$ and the low-mass spectral shape.

\FloatBarrier %

\end{document}